\documentclass[twocolumn,english,superscriptaddress]{revtex4-1}
\usepackage{graphicx}
\usepackage[english]{babel}

\begin{document}

\title{Universal Set of Quantum Gates for Double-Dot Exchange-Only Spin Qubits with Intradot Coupling}

\author{Marco De Michielis}
\email{marco.demichielis@mdm.imm.cnr.it}
\affiliation{Laboratorio MDM, Istituto per la Microelettronica e Microsistemi, Consiglio Nazionale delle Ricerche, Via C. Olivetti 2, I-20864 Agrate Brianza (MB), Italy}

\author{Elena Ferraro}
\affiliation{Laboratorio MDM, Istituto per la Microelettronica e Microsistemi, Consiglio Nazionale delle Ricerche, Via C. Olivetti 2, I-20864 Agrate Brianza (MB), Italy}

\author{Marco Fanciulli}
\affiliation{Laboratorio MDM, Istituto per la Microelettronica e Microsistemi, Consiglio Nazionale delle Ricerche, Via C. Olivetti 2, I-20864 Agrate Brianza (MB), Italy}
\affiliation{Dipartimento di Scienza dei Materiali, Universit\`a degli Studi Milano-Bicocca, Via Cozzi 53, I-20125 Milano, Italy}

\author{Enrico Prati}
\affiliation{Laboratorio MDM, Istituto per la Microelettronica e Microsistemi, Consiglio Nazionale delle Ricerche, Via C. Olivetti 2, I-20864 Agrate Brianza (MB), Italy}
\affiliation{Istituto di Fotonica e Nanotecnologia, Consiglio Nazionale delle Ricerche, Piazza Leonardo da Vinci 32, I-20133 Milano, Italy}

\begin{abstract}
We present a universal set of quantum gate operations based on exchange-only spin qubits in a double quantum dot, where each qubit is obtained by three electrons in the (2,1) filling.
Gate operations are addressed by modulating electrostatically the tunneling barrier and the energy offset between the two dots, singly and doubly occupied respectively. 
We propose explicit gate sequences of single qubit operations for Hadamard gate and $\pi$/8 gate, and the two-qubit controlled NOT (CNOT) gate, to complete the universal set. 
The unswitchable interaction between the two electrons of the doubly occupied quantum dot is taken into account. Short gate times are obtained by employing spin density functional theory simulations.
\end{abstract}

\maketitle

Electron spins confined in semiconductor quantum dots (QDs) have been employed to implement basis of increasingly complex angular momentum for quantum computing \cite{Koppens_Nature2006,Morello_Nature10, HRL_Nature2012, Medford_NatNano2013}. In the simplest proposals, a single electron spin forms the logical basis for single qubit operations performed via spin resonance \cite{Loss_PRA1998,Vrijen_PRA2000}. 
An alternative scheme, with logical basis formed from singlet and triplet states of two electron spins which immunizes qubits against the dominant error from hyperfine interactions \cite{Levy_PRL2002,Petta_Science2005,Taylor_NatPhys2005}, requires inhomogeneous static magnetic field for full single-qubit control. Exchange-interaction only qubits based on three electron spins removes the need for an inhomogeneous field as interactions between adjacent electron spins suffice for all one and two qubit operations \cite{DiVincenzo_Nature2000,Medford_NatNano2013}.
A compact variant of the three spin qubit in (1,1,1) states of three QDs proposed by DiVincenzo \cite{DiVincenzo_Nature2000} has been recently developed for (2,1) states in double QDs, with total spin states belonging to the subspace $S$=1/2 and $S_z$=-1/2 \cite{Shi_PRL2012,Koh_PRL2012,Ferraro_QIP2014,Shi_NatureComm2014}. Differently from the (1,1,1) system, the (2,1) system offers the advantages of higher protection from hyperfine interactions of the singlet and triplet state in one of the two dots \cite{Levy_PRL2002}, and compact fabrication - only two dots instead of three \cite{Prati_Nano2012,Pierre09}. 
A possible drawback consists of more constrained interactions, as it is not possible to tune separately the exchange of the two electrons sharing the same site with the third electron spin. In addition, exchange
interaction between the electrons in the doubly occupied dot cannot be effectively turned off. While a case limited to the first aspect has been considered in the past \cite{Shi_PRL2012}, the demonstration of a universal set of quantum logic gates when the inescapable intra-dot interaction of a realistic QD is included, still lacks.
The single qubit operations proposed in \cite{Shi_PRL2012} for extending the ports are incorrect. 
Universal quantum computation in an alternative scheme, where spin and valley degrees of freedom of two electrons in two QDs are exploited, has been proposed \cite{Rohling_NJP2012}.

Here, we report on a universal set of gates for two exchange-only qubits in double QDs with fixed intradot exchange, generated by a Hadamard gate, a $\pi$/8 gate (single qubit operations), and a controlled-NOT (CNOT) gate based on two qubits. For each gate interaction, sequences supplied with operation times are calculated by employing a genetic algorithms which takes into account the intradot interaction of the electrons in the doubly occupied QD.
One and two qubits systems are modeled by effective Hamiltonians and a dynamical evolution operator is used to find the overall effect of the interactions. We developed and used a search algorithm to find the interactions sequences. A Spin Density Functional Theory (SDFT) simulator is used to estimate the parameters of the effective Hamiltonians, revealing short gate times for high performances.

\section{The Hybrid Qubit}
\begin{figure}[h]
\includegraphics[width=0.45\textwidth]{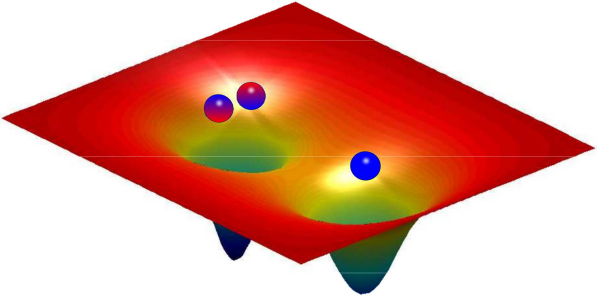}
\caption{\label{Fig:HDQDqubit} Pictorial representation of the hybrid double quantum dot qubit with three electrostatically confined electrons. The quantum state $ \left| S \right \rangle \left| \downarrow \right\rangle $ is pictured. }
\end{figure}

The hybrid qubit, whose energy landscape is shown in Fig.\ref{Fig:HDQDqubit}, is composed of a double QD with two electrons in one QD (left) and one electron in the other QD (right).
Let's now define the logical basis, enumerating the possible
transitions between the three electrons spin states that can be induced
by manipulations which preserve total spin angular momentum. The Hilbert
space of three electron spins has eight possible spin states: the total
spin eigenstates form indeed a quadruplet with $S = 3/2$ and $S_z = \pm
3/2, \pm 1/2$ and two doublets each with $S = 1/2$ and $S_z = \pm 1/2$, where
the square of the total spin is $\hbar^2 S(S + 1)$ and the $z$-component of
the total spin is $\hbar S_z$.
The logical qubit space is chosen to be in the two-dimensional subspace
with $S=1/2$, $S_z=-1/2$. We point out that only states with the same $S$ and
$S_z$ can be coupled by spin independent terms in the Hamiltonian.
The value of the total angular momentum operator $S$ specifies whether the
decoherence free subsystem qubit has leaked; $S=1/2$ is unleaked while
$S=3/2$ is leaked.
The logical basis $\{|0\rangle, |1\rangle\}$ written via Clebsch-Gordan
coefficients is given by
\begin{equation}
\left| 0 \right \rangle \equiv \left| S \right \rangle \left| \downarrow \right\rangle,  \; \; \; \; \; \;
\left| 1 \right \rangle \equiv  \sqrt{\frac{1}{3}}\left| T_{0} \right \rangle \left| \downarrow \right\rangle - \sqrt{\frac{2}{3}}\left| T_{-} \right \rangle \left| \uparrow \right\rangle 
\end{equation}
where $\left| S \right \rangle$, $\left| T_{0} \right \rangle$ and $\left| T_{-} \right \rangle $ are respectively the singlet and triplet states of the pair in the left dot, in combination with the angular momentum state of the electron spin localized in the right dot. 

The hybrid qubit is described by an Hubbard-like model that, following the Schieffer-Wolff projection operator method, could be recast in a spin Hamiltonian \cite{Ferraro_QIP2014}. The effective Hamiltonian is expressed as a sum of exchange interactions between each pairs of electron spins
\begin{equation} \label{Eq:singleQubit}
H^{eff} \approx J_{13} \mathbf{S}_{1} \cdot \mathbf{S}_{3} + J_{23} \mathbf{S}_{2} \cdot \mathbf{S}_{3} + J_{12} \mathbf{S}_{1} \cdot \mathbf{S}_{2}
\end{equation}
with exchange interactions
\begin{eqnarray} \label{Eq:Js}
J_{13} &\simeq&  \frac{1}{E_{(012)} - E_{(111)}} 4(t_{13}-J_{t}^{13})^2 - 2J_{e}^{13} \nonumber \\
J_{23} &\simeq&  \frac{1}{E_{(102)} - E_{(111)}} 4(t_{23}-J_{t}^{23})^2 - 2J_{e}^{23} \nonumber \\
J_{12} &\simeq&  \left ( \frac{1}{E_{(201)} - E_{(111)}}+\frac{1}{E_{(021)}-E_{(111)}} \right ) 4(J_{t}^{12})^2 - 2J_{e}^{12}, \nonumber \\
\end{eqnarray}
where $E_{(\alpha \beta \gamma)}$ are the energies with $\alpha$ ($\beta$) electrons in the ground state (first excited) of the left dot and $\gamma$ electrons in the second dot. The parameters $t_{ij}$ are the tunneling rates, $J_{t}^{ij}$ account for the occupation-modulated hoppings, $J_{e}$ are the spin-exchange terms from the energy level $i$ to $j$ as defined in \cite{Jefferson_PRB1996}.

Differently from inter-QD interactions governed by tunable $J_{13}$ and $J_{23}$, the intra-QD interaction $J_{12}$ can not be effectively controlled. In fact $J_{12}$ does not depend on the tunneling rates (see Eq. (\ref{Eq:Js})) whose values can span several order of magnitudes and strongly control $J_{13}$ and $J_{23}$.
We assumed $\max(J_{12})=\max(J_{13})=J^{max}$ and we set a realistic value for $J_{12}=J^{max}/2$ to model the control ineffectiveness. The exchange interactions $J_{13}(t)$ and $J_{23}(t)$ are assumed to have instantaneous turn-on and turn-off as in \cite{Shi_PRL2012}. 

\begin{figure}[h]
\includegraphics[width=0.49\textwidth]{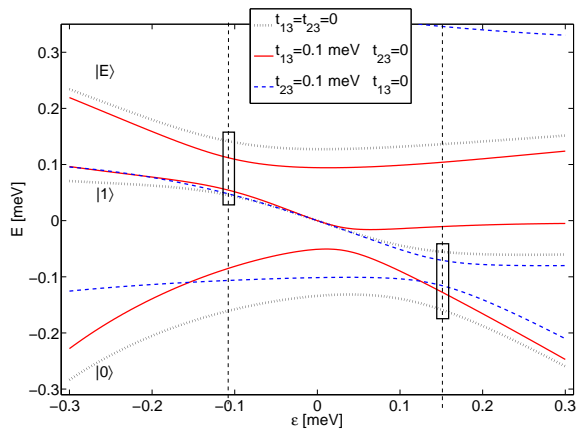}
\caption{\label{Fig:eigenval_t13_t23} Energy levels of Hamiltonian (\ref{effmatrix2}) as a function of detuning $\varepsilon$ for three different configurations of the tunneling rates between the two dots: energy levels of the system when $t_{13}$=$t_{23}$=0 are highlighted in dotted black lines, $t_{13}$ on and $t_{23}$ off in solid red lines and $t_{23}$ on and $t_{13}$ off in broken blue curves. The boxes highlight where the system has to be biased to maximize $J_{13}$ (on the left) and $J_{23}$ (on the right), with $\max(J_{13})=\max(J_{23}) \equiv J^{max}$.}
\end{figure}

In order to clarify how tunneling rates $t_{ij}$ and the energy detuning $\varepsilon$ between the two QDs can control the hybrid qubit, a simplified Hamiltonian for an hybrid qubit as a function of the inter-QD tunneling rates and of the inter-QD energy detuning is derived, recovering similar results obtained with the heuristic Hamiltonian in Ref. \cite{Koh_PRL2012}.
We consider a basis with an intermediated state $|E\rangle\equiv|\downarrow\rangle|S\rangle$ in addition to the logical states $|0\rangle$ and $|1\rangle$. The state $|E\rangle$, which has one electron in the left dot and two electrons in the right dot conserving the same total angular momentum $S^2$ and $S_z$, is directly involved in the physical process that leads to transitions between the two logical states.
Eq. \ref{effmatrix2} reports the effective Hamiltonian in the basis $\{|0\rangle,|1\rangle,|E\rangle\}$ where the inter-QD detuning $\varepsilon$ is introduced.
\begin{widetext}
\begin{equation}\label{effmatrix2}
H^{3 \times 3} =
\left( 
\begin{array}{ccc}
-\frac{3}{4}J_{12} & -\frac{\sqrt{3}}{4}(J_{13}-J_{23}) & \frac{3}{8}(J_{23}-J_{13}+J_{12}) \\
-\frac{\sqrt{3}}{4}(J_{13}-J_{23}) & \frac{1}{4}J_{12}-\frac{1}{2}(J_{13}+J_{23}) & -\frac{\sqrt{3}}{8}(J_{13}+3J_{23}-J_{12})\\
\frac{3}{8}(J_{23}-J_{13}+J_{12}) & -\frac{\sqrt{3}}{8}(J_{13}+3J_{23}-J_{12}) & -\frac{3}{4}J_{23}-\varepsilon
\end{array} 
\right)
\end{equation}
\end{widetext}
$H^{3\times3}$ eigenvalues are reported in Fig. \ref{Fig:eigenval_t13_t23} as a function of $\varepsilon$ in three different cases: both tunneling rates are zero; $t_{13}$ is on and $t_{23}$=0 and $t_{23}$ is on and $t_{13}$=0. 
Transitions from logical state $|0\rangle$ to $|1\rangle$ can be induced by first setting $\varepsilon$ to the avoided crossing between $|0\rangle$ and $|E\rangle$ when $t_{23}$ is switched on (dashed blue curves in the right box) and then switching $t_{23}$ on and off. Then, changing $\varepsilon$ to the avoided crossing between $|E\rangle$ and $|1\rangle$ when $t_{13}$ is on (solid red lines in the left box) and pulsing $t_{13}$ on and off. 
The same argument can be applied to induce transition conversely from logical state $|1\rangle$ to $|0\rangle$.

In order to obtain the exchange interaction sequences for the three different quantum gates, we developed a search algorithm similar to the one described in Ref. \cite{Fong_QIC2011}, which is a combination of a simplex-based and a genetic algorithms. At each iteration of the search algorithm sequences become closer to the global minimum, featuring a reduced number of exchange steps and minimum interaction time per step.
The qubit sequences are calculated under the assumption that a sufficiently high number of external inputs (electrostatic gates) are available to finely control the band structure of the double QDs.

\section{Gate Sequences for Hadamard and  $\pi$/8 Gates}
\begin{figure}[h]
a)
\begin{equation}
H=\frac{1}{\sqrt{2}}\left(\begin{array}{cc} 
1 & 1 \\
1 & -1 
\end{array}\right)
\end{equation}
\begin{center}
\includegraphics[clip=true, width=0.4\textwidth]{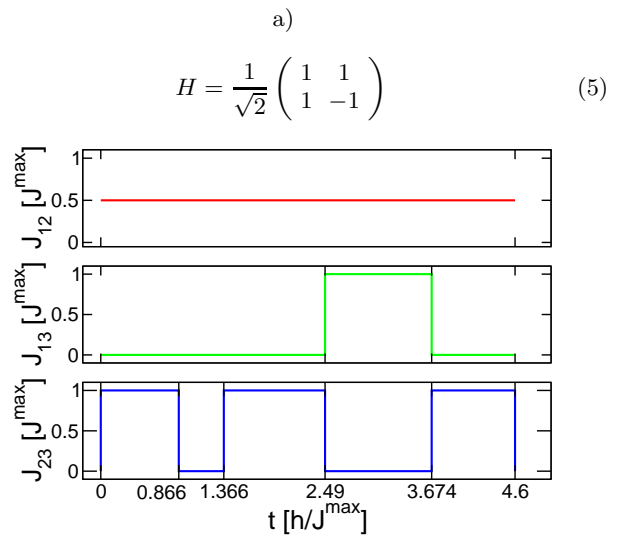}
\end{center}
b)
\begin{equation}
R_{\pi/4}=\left(\begin{array}{cc} 
1 & 0  \\
0 & e^{i\pi/4} 
\end{array}\right)
\end{equation}
\centering
\includegraphics[clip=true, width=0.4\textwidth]{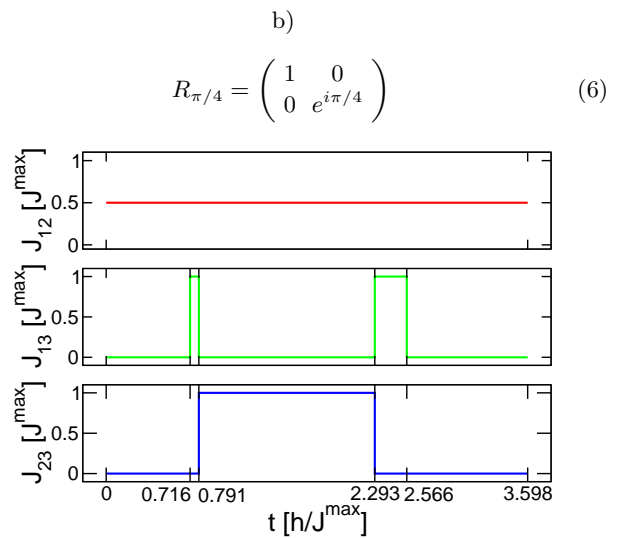}
\caption{\label{Fig:SingleQubitOp} Matrix representation and exchange sequence implementing a) Hadamard and b) $\pi/8$ gates up to a global phase. $J_{12}$, $J_{13}$ and $J_{23}$ are plotted as a function of the time $t$ expressed in units of $h/J^{max}$. Note that the constant interaction $J_{12}$=$J^{max}$/2.}
\end{figure}

Single qubit operation sequences are reported in the following. In Fig.\ref{Fig:SingleQubitOp} we recall the matrix representation and present the interaction sequences for the Hadamard and $\pi$/8 gates up to a global phase, respectively, calculated by using our search algorithm. Both gate operations are implemented by sequences of five steps.

\section{Gate Sequences for an Exact CNOT} 
\begin{figure}[h]
\centering
\includegraphics[width=0.3\textwidth]{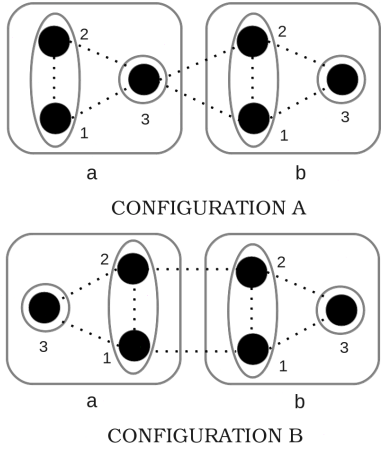}
\caption{\label{Fig:confs} Couple of hybrid qubits: configurations A and B. Configurations differ for the orientation of qubit a and, consequentially, for the inter-qubit connections.}
\end{figure}

When considering two qubits operations, the two most significant configurations, where the number of inter-qubit connections is maximized, are reported in Fig.\ref{Fig:confs}. 

Dropping the \textit{eff} superscripts, the effective Hamiltonian of the couple of hybrid qubit in the $K$ configuration is: 
\label{Eq:hamiltonianCouple}
\begin{equation}
H^{K}_{ab}=H_a+H_b+H^{K}_{int}
\end{equation}
(for K=A,B) where $H_a$, $H_b$ are the effective Hamiltonians of the single qubits a and b, respectively (see Eq. (\ref{Eq:singleQubit})), and $H_{int}^{K}$ is the interaction Hamiltonian. 
When configuration A is considered, 
\begin{equation} \label{Eq:InterQubitHamiltonianA}
H^{A}_{int}=\sum_{i=1}^{2}J_{3_ai_b} \mathbf{S}_{3_a}\cdot\mathbf{S}_{i_b},
\end{equation}
whereas when $K$=$B$
\begin{equation} \label{Eq:InterQubitHamiltonianB}
H^{B}_{int}=\sum_{i=1}^{2}J_{i_ai_b} \mathbf{S}_{i_a}\cdot\mathbf{S}_{i_b}.
\end{equation}

\begin{figure}[h!]
\begin{equation}
CNOT=\left(\begin{array}{cccc} 
1 & 0 & 0 & 0\\
0 & 1 & 0 & 0\\
0 & 0 & 0 & 1\\
0 & 0 & 1 & 0
\end{array}\right)
\end{equation}
\vspace{-0.5cm}
\begin{center}
\includegraphics[clip=true, width=0.39\textwidth]{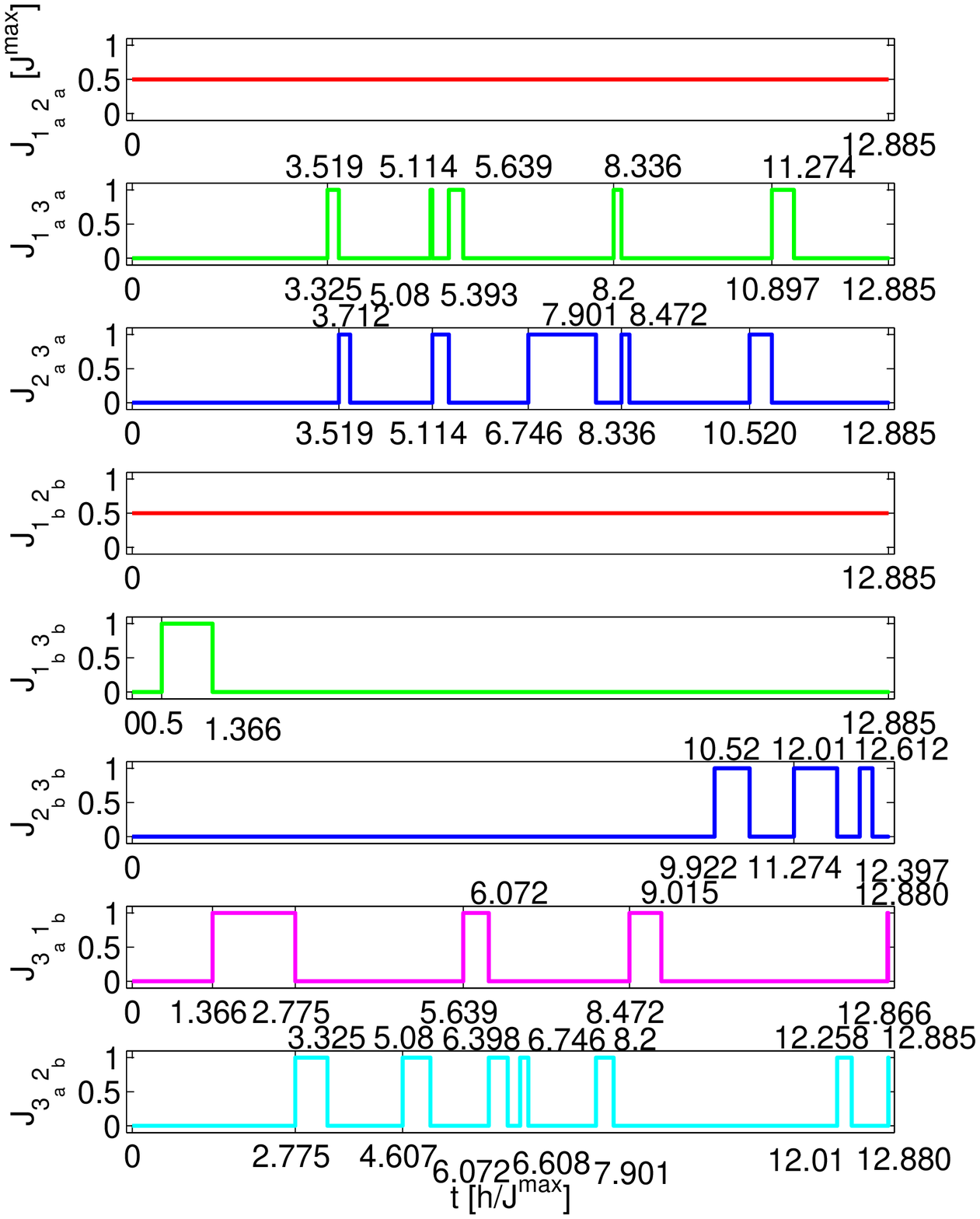}
\end{center}
\vspace{-1cm}
\centering
Configuration A

\vspace{-0.3cm}
\begin{center}
\includegraphics[clip=true, width=0.39\textwidth]{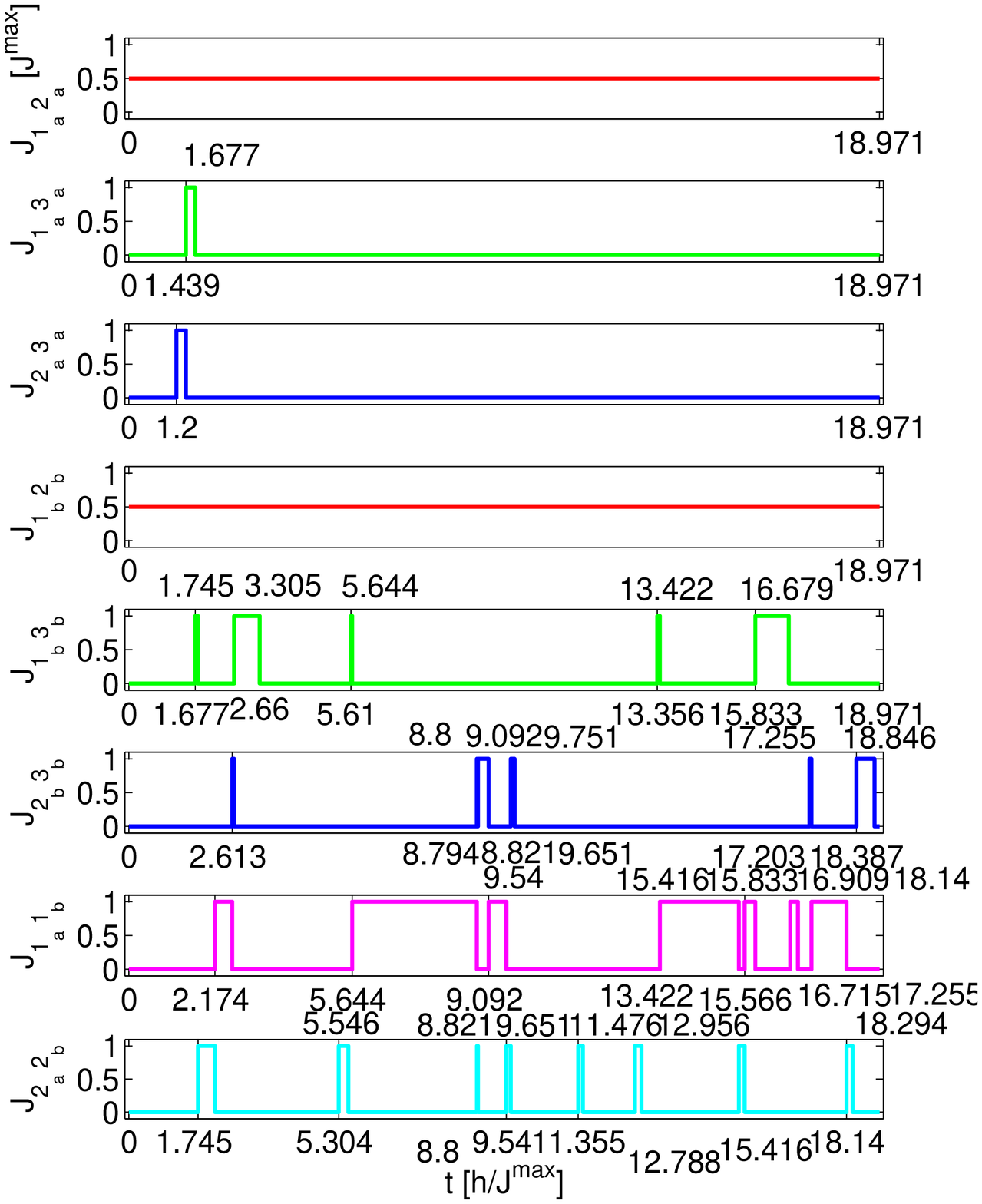}
\end{center}
\vspace{-1cm}
\centering
Configuration B

\caption{\label{Fig:SeqsCNOTs} Matrix representation and exchange sequence  for an exact CNOT operation (up to a global phase) with fixed $J_{1_a 2_a}=J_{1_b 2_b}=J^{max}$/2 as a function of time $t$ for both configurations A and B.}
\end{figure}

The detailed derivation of the two qubit Hamiltonians is reported in \cite{SupMatDeMichielis_EPL2014}. 
The qubit gate sequences for exact CNOT gates with fixed $J_{12}$=$J^{max}$/2 in both qubits for the configuration A and B are presented in Fig. \ref{Fig:SeqsCNOTs}. 
These sequences differs from those reported in Ref. \cite{Shi_PRL2012} because ours account for ineffectively controlled interactions $J_{12}$ and because we provide exacts CNOTs instead of locally equivalent ones only. Note that the CNOT sequence for configuration B reported in Ref. \cite{Shi_PRL2012} provides only a correct locally equivalent CNOT because the corresponding single qubit operations needed to transform a locally equivalent CNOT to an exact CNOT are incorrect.

\section{Gate Performances}
Sequence times are evaluated for each gate by estimating $J^{max}$ through a simulator based on Spin Density Function Theory (SDFT) \cite{VonBarth_JPC1972}. 
For each couple of valleys of silicon along $\Delta$ crystallographic directions, the simulator solves the Kohn-Sham equations in the Effective Mass Approximation (EMA) with anisotropic effective masses for both spin down and spin up populations \cite{DeMichielis_APEX2012}. When the eigenstates are obtained, the spin density concentrations are calculated and the effective potentials, namely the Hartree and the exchange-correlation potentials, are derived. For the exchange-correlation potential a Local Density Approximation (LDA) is assumed by using the parametrization presented in \cite{Perdew_PRB1992}.
The total potential is then calculated self-consistently by solving the Poisson equation with the applied potentials from the external. The simulation ends when the error between the potential of the current iteration and that of the previous one is under a given tolerance.

\begin{figure}[ht]
\begin{center}
\includegraphics[width=0.4\textwidth]{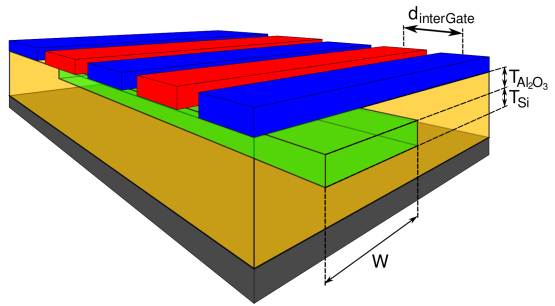}
\end{center}
\caption{The single qubit device is modeled as a silicon nanowire (in green) embedded in an insulator slab (in yellow). Accumulation gates forming the QDs are highlighted in red whereas the contacts controlling the inter-QD electrostatic barriers are shown in blue. The back gate is shown in gray.}\label{Fig:deviceModel} 
\end{figure}

To exemplify a realistic condition, we consider a double QD created in a Si nanowire featuring a rectangular section with a thickness $T_{Si}$=15 nm and  width $W$=60 nm on a thick layer of SiO$_{2}$ as pictured in Fig. \ref{Fig:deviceModel}. An Al$_{2}$O$_{3}$ layer with thickness $T_{Al_2O_3}$=40 nm is deposed on the nanowire and Al gates placed orthogonally to nanowire direction and separated by $d_{interGate}$ are used to electrostatically confine electrons and control the inter-QD tunneling rates in the underneath silicon. 
In order to exploit quantum states from a single $\Delta$ valley, the valley splitting $\Delta E_{v}$ is enhanced to $\Delta E_{v}$$\sim$ 500 $\mu$eV by increasing the electric field at the Si/Al$_{2}$O$_{3}$ interface by polarizing negatively the back gate.  
Simulations show that, when $d_{interQD} \equiv 2d_{interGate}$=40 nm, the maximum effective interaction $J^{max}$= 7.2 $\mu$eV, providing operation times of $t_{H}$=2.64 ns and $t_{\pi/8}$= 2.07 ns for the Hadamard and $\pi$/8 gates, respectively. Calculated CNOT gate times are $t^{A}_{CNOT}$=7.40 ns and $t^{B}_{CNOT}$=10.9 ns for two qubit configuration A and B, respectively. 
Effects of charge and spin noise on the hybrid qubit decoherence in natural silicon with DC pulsed gating scheme are reported in Ref. \cite{Koh_PNAS2013} where a fidelity of 99.995\% for the z-rotations and 83\% for the x-rotation have been estimated.

\section{Summary} 
We have presented a universal set of quantum gates for hybrid double quantum dot qubits with realistic non-vanishing intradot interaction, composed by Hadamard, $\pi$/8 and CNOT gates. 
By using a versatile search algorithm, feasible interaction sequences have been reported for each gate by taking into account not switchable interactions between the two electrons confined inside the same quantum dot.
The two principal configurations have been studied by coupling two qubits. Under such conditions, we obtain the exact CNOT sequences for realistic hybrid qubits in both configurations, including single qubit operations.
CNOT gate operations times in the range of 10 ns are predicted by employing SDFT simulations.

\bibliography{biblio.bib}

\end{document}


\title{Universal Set of Quantum Gates for Double-Dot Exchange-Only Spin Qubits with Intradot Coupling: Supplemental Material}

\author{Marco De Michielis}
\email{marco.demichielis@mdm.imm.cnr.it}
\affiliation{Laboratorio MDM, Istituto per la Microelettronica e Microsistemi, Consiglio Nazionale delle Ricerche, Via C. Olivetti 2, I-20864 Agrate Brianza (MB), Italy}

\author{Elena Ferraro}
\affiliation{Laboratorio MDM, Istituto per la Microelettronica e Microsistemi, Consiglio Nazionale delle Ricerche, Via C. Olivetti 2, I-20864 Agrate Brianza (MB), Italy}

\author{Marco Fanciulli}
\affiliation{Laboratorio MDM, Istituto per la Microelettronica e Microsistemi, Consiglio Nazionale delle Ricerche, Via C. Olivetti 2, I-20864 Agrate Brianza (MB), Italy}
\affiliation{Dipartimento di Scienza dei Materiali, Universit\`a degli Studi Milano-Bicocca, Via Cozzi 53, I-20125 Milano, Italy}

\author{Enrico Prati}
\affiliation{Laboratorio MDM, Istituto per la Microelettronica e Microsistemi, Consiglio Nazionale delle Ricerche, Via C. Olivetti 2, I-20864 Agrate Brianza (MB), Italy}
\affiliation{Istituto di Fotonica e Nanotecnologia, Consiglio Nazionale delle Ricerche, Piazza Leonardo da Vinci 32, I-20133 Milano, Italy}

\begin{abstract}
This supplementary material presents the derivation of the expressions of the effective Hamiltonian models of two interacting hybrid qubits reported in the main text.
\end{abstract}

\maketitle

We present the explicit expressions of the effective Hamiltonian models of two interacting hybrid qubits. The total Hamiltonian model is composed by two free terms describing the two qubits and by an interaction part between the two.

The free Hamiltonians of the two qubits $a$ and $b$, written in terms of the creation and annihilation fermionic operators $c_k^{\dagger}$ and $c_k$ respectively are given for $q\equiv a,\,b$ by
\begin{equation}
H_{q}=H_{e_q}+H_{t_q}+H_{U_q}+H_{J_q},
\end{equation}
where terms on the right side are reported in Eqs. (\ref{Eq:Heq} - \ref{hj}) for every pair of spins considered.
\begin{eqnarray} \label{Eq:Heq}
H_{e_q}&=&\sum_{k=1_q,\sigma}^{3_q}\varepsilon_kc^{\dagger}_{k\sigma}c_{k\sigma}
\end{eqnarray}
\begin{eqnarray} \label{Eq:Htq}
H_{t_q}&=&t_{1_q3_q}\sum_{\sigma}(c^{\dagger}_{1_q\sigma}c_{3_q\sigma}+h.c.)+t_{2_q3_q}\sum_{\sigma}(c^{\dagger}_{2_q\sigma}c_{3_q\sigma}+h.c.)
\end{eqnarray}
\begin{eqnarray} \label{Eq:HUq}
H_{U_q}&=&\sum_{k=1_q}^{3_q}U_kn_{k\uparrow}n_{k\downarrow}+U_{1_q2_q}(n_{1_q\uparrow}+n_{1_q\downarrow})(n_{2_q\uparrow}+n_{2_q\downarrow})+\nonumber\\ &+&U_{1_q3_q}(n_{1_q\uparrow}+n_{1_q\downarrow})(n_{3_q\uparrow}+n_{3_q\downarrow})+\nonumber \\
&+& U_{2_q3_q}(n_{2_q\uparrow}+n_{2_q\downarrow})(n_{3_q\uparrow}+n_{3_q\downarrow})
\end{eqnarray}
\begin{eqnarray} \label{Eq:HJq}
H_{J_q}&=& H^{(1_q3_q)}_J+H^{(2_q3_q)}_J+H^{(1_q2_q)}_J  
\end{eqnarray}
\begin{eqnarray}\label{hj}
H^{(i_qj_q)}_J&=&-J^{(i_qj_q)}_e(n_{i_q\uparrow}n_{j_q\uparrow}+n_{i_q\downarrow}n_{j_q\downarrow})-( J^{(i_qj_q)}_ec^{\dagger}_{i_q\downarrow}c^{\dagger}_{j_q\uparrow}c_{j_q\downarrow}c_{i_q\uparrow}+
J^{(i_qj_q)}_pc^{\dagger}_{j_q\uparrow}c^{\dagger}_{j_q\downarrow}c_{i_q\uparrow}c_{i_q\downarrow}+ \nonumber \\ &+&\sum_{k_q,\sigma}J^{(i_qj_q)}_tn_{k_q\sigma}c^{\dagger}_{i_q\bar{\sigma}}c_{j_q\bar{\sigma}}+h.c.). 
\end{eqnarray}
$H_{e_q}$ and $H_{t_q}$ describe respectively the single electron energy level of each dot and the tunneling energy. The last two terms $H_{U_q}$ and $H_{J_q}$ constitute the intra-dot and inter-dot Coulomb interactions.  $J^{(i_qj_q)}_e$ is the spin exchange term, $J^{(i_qj_q)}_p$ the pair-hopping term and $J^{(i_qj_q)}_t$ the occupation-modulated one.

Regarding the configuration A the interaction Hamiltonian is given by the sum of the following terms
\begin{eqnarray}
H_t &=& t_{3_a1_b}\sum_{\sigma}(c^{\dagger}_{3_a\sigma}c_{1_b\sigma}+h.c.)+t_{3_a2_b}\sum_{\sigma}(c^{\dagger}_{3_a\sigma}c_{2_b\sigma}+h.c.)\nonumber\\
H_U &=& U_{3_a1_b}(n_{3_a\uparrow}+n_{3_a\downarrow})(n_{1_b\uparrow}+n_{1_b\downarrow})+ \nonumber \\ &+& U_{3_a2_b}(n_{3_a\uparrow}+n_{3_a\downarrow})(n_{2_b\uparrow}+n_{2_b\downarrow})\nonumber\\
H_J &=& H^{(3_a1_b)}_J+H^{(3_a2_b)}_J,\nonumber
\end{eqnarray}
where $H_J$ is defined as in Eq.(\ref{hj}).

Following the same procedure reported in Ref. \cite{Ferraro_QIP2014}, we calculate the projected Hamiltonian using an appropriate operator. The effective Hamiltonian, concerning the low energy excitation, appears as the sum of all the exchange interactions between pairs of spins
\begin{eqnarray}\label{Heff2A}
H^{eff}&=&\sum_{q=a,b}(J_{1_q3_q}\mathbf{S}_{1_q}\cdot\mathbf{S}_{3_q}+J_{2_q3_q}\mathbf{S}_{2_q}\cdot\mathbf{S}_{3_q}+J_{1_q2_q}\mathbf{S}_{1_q}\cdot\mathbf{S}_{2_q})+ \nonumber \\
&+& J_{3_a1_b}\mathbf{S}_{3_a}\cdot\mathbf{S}_{1_b}+J_{3_a2_b}\mathbf{S}_{3_a}\cdot\mathbf{S}_{2_b}.
\end{eqnarray}
The effective coupling constants, considering that intra-dot Coulomb energies are larger than all the other contributions are finally given by
\begin{eqnarray}\label{couplingA}
J_{1_q3_q}&\simeq&\frac{1}{\Delta E_{1_q}}4(t_{1_q3_q}-J^{(1_q3_q)}_t)^2-2J^{(1_q3_q)}_e\nonumber\\
J_{2_q3_q}&\simeq&\frac{1}{\Delta E_{2_q}}4(t_{2_q3_q}-J^{(2_q3_q)}_t)^2-2J^{(2_q3_q)}_e\nonumber\\
J_{1_q2_q}&=&\left(\frac{1}{\Delta E_{3_q}}+\frac{1}{\Delta E_{4_q}}\right)4J^{(1_q2_q)2}_t-2J^{(1_q2_q)}_e\nonumber\\
J_{3_a1_b}&\simeq&\frac{1}{\Delta E_5}4(t_{3_a1_b}-J^{(3_a1_b)}_t)^2-2J^{(3_a1_b)}_e\nonumber\\
J_{3_a2_b}&\simeq&\frac{1}{\Delta E_6}4(t_{3_a2_b}-J^{(3_a2_b)}_t)^2-2J^{(3_a2_b)}_e,\nonumber
\end{eqnarray}
where the energy differences for qubits $a$ and $b$ are 
\begin{eqnarray}\label{deltaa}
\Delta E_{1_{a(b)}}&=& E_{(012,111)}(E_{(111,012)})-E_{(111,111)}\nonumber\\
\Delta E_{2_{a(b)}}&=& E_{(102,111)}(E_{(111,102)})-E_{(111,111)}\nonumber\\
\Delta E_{3_{a(b)}}&=& E_{(201,111)}(E_{(111,201)})-E_{(111,111)}\nonumber\\
\Delta E_{4_{a(b)}}&=& E_{(021,111)}(E_{(111,021)})-E_{(111,111)}\nonumber\\
\Delta E_5&=& E_{(112,011)}-E_{(111,111)}\nonumber\\
\Delta E_6&=& E_{(112,101)}-E_{(111,111)} \nonumber
\end{eqnarray}
with the different terms defined in Eqs. \ref{a1}, \ref{a2}, \ref{a3} where 
we point out that the first (last) three indices inside parenthesis, $0\le i\neq j\neq k\le 2$, assuming only integer values, denote the number of electrons in each level for qubit a (b).
\begin{eqnarray}\label{a1}
E_{(ijk,111)}&=& i\varepsilon_{1_a}+j\varepsilon_{2_a}+k\varepsilon_{3_a}+ijU_{1_a2_a}+ikU_{1_a3_a}+kjU_{2_a3_a}+\delta_{i2}U_{1_a}+\delta_{j2}U_{2_a}+\nonumber\\
&+& \delta_{k2}U_{3_a}+\varepsilon_{1_b}+\varepsilon_{2_b}+\varepsilon_{3_b}+U_{1_b2_b}+U_{1_b3_b}+U_{2_b3_b}+kU_{3_a1_b}+kU_{3_a2_b}
\end{eqnarray}
\begin{eqnarray}\label{a2}
E_{(111,ijk)}&=&\varepsilon_{1_a}+\varepsilon_{2_a}+\varepsilon_{3_a}+U_{1_a2_a}+U_{1_a3_a}+U_{2_a3_a}+i\varepsilon_{1_b}+j\varepsilon_{2_b}+k\varepsilon_{3_b}+ijU_{1_b2_b}+\nonumber\\
&+& ikU_{1_b3_b}+kjU_{2_b3_b}+\delta_{i2}U_{1_b}+\delta_{j2}U_{2_b}+\delta_{k2}U_{3_b}+iU_{3_a1_b}+iU_{3_a2_b}
\end{eqnarray}
\begin{eqnarray}\label{a3}
E_{(111,111)}&=&\varepsilon_{1_a}+\varepsilon_{2_a}+\varepsilon_{3_a}+U_{1_a2_a}+U_{1_a3_a}+U_{2_a3_a}+\varepsilon_{1_b}+\varepsilon_{2_b}+\varepsilon_{3_b}+\nonumber\\
&+& U_{1_b2_b} + U_{1_b3_b}+U_{2_b3_b}+U_{3_a1_b}+U_{3_a2_b}
\end{eqnarray}

For the second configuration B under study we assume that the energy detuning between the double occupied QDs is small enough so terms of the Hamiltonian containing tunnelling rates and exchange interactions between energy levels in different QDs with different indexes (i.e. $1_a2_b$, $2_a1_b$) are negligible. The interaction Hamiltonian terms are finally given by
\begin{eqnarray}
H_t&=& t_{1_a1_b}\sum_{\sigma}(c^{\dagger}_{1_a\sigma}c_{1_b\sigma}+h.c.)+ t_{2_a2_b}\sum_{\sigma}(c^{\dagger}_{2_a\sigma}c_{2_b\sigma}+h.c.) \nonumber\\
H_U &=& U_{1_a1_b}(n_{1_a\uparrow}+n_{1_a\downarrow})(n_{1_b\uparrow}+n_{1_b\downarrow})+ \nonumber \\
&+& U_{1_a2_b}(n_{1_a\uparrow}+n_{1_a\downarrow})(n_{2_b\uparrow}+n_{2_b\downarrow})+\nonumber\\
&+& U_{2_a1_b}(n_{2_a\uparrow}+n_{2_a\downarrow})(n_{1_b\uparrow}+n_{1_b\downarrow})+ \nonumber \\
&+&U_{2_a2_b}(n_{2_a\uparrow}+n_{2_a\downarrow})(n_{2_b\uparrow}+n_{2_b\downarrow})\nonumber\\
H_J&=& H^{(1_a1_b)}_J+H^{(2_a2_b)}_J \nonumber
\end{eqnarray}
where $H_J$ is defined as in Eq.(\ref{hj}). 

Analogously to the previous case, the effective Hamiltonian appears as the sum of exchange interactions:
\begin{eqnarray}\label{Heff2B}
H^{eff}&=&\sum_{q=a,b}(J_{1_q3_q}\mathbf{S}_{1_q}\cdot\mathbf{S}_{3_q}+J_{2_q3_q}\mathbf{S}_{2_q}\cdot\mathbf{S}_{3_q} + J_{1_q2_q}\mathbf{S}_{1_q}\cdot\mathbf{S}_{2_q})+\nonumber\\
&+& J_{1_a1_b}\mathbf{S}_{1_a}\cdot\mathbf{S}_{1_b}+J_{1_a2_b}\mathbf{S}_{1_a}\cdot\mathbf{S}_{2_b}+ \nonumber \\
&+& J_{2_a1_b}\mathbf{S}_{2_a}\cdot\mathbf{S}_{1_b}+J_{2_a2_b}\mathbf{S}_{2_a}\cdot\mathbf{S}_{2_b}.
\end{eqnarray}
The effective coupling constants, under the assumption of larger intra-dot energies with respect to inter-dot ones are finally given by
\begin{eqnarray}\label{couplingB}
J_{1_q3_q}&\simeq&\frac{1}{\Delta E_{1_q}}4(t_{1_q3_q}-J^{(1_q3_q)}_t)^2-2J^{(1_q3_q)}_e\nonumber\\
J_{2_q3_q}&\simeq&\frac{1}{\Delta E_{2_q}}4(t_{2_q3_q}-J^{(2_q3_q)}_t)^2-2J^{(2_q3_q)}_e\nonumber\\
J_{1_q2_q}&=&\left(\frac{1}{\Delta E_{3_q}}+\frac{1}{\Delta E_{4_q}}\right)4J^{(1_q2_q)2}_t-2J^{(1_q2_q)}_e\nonumber\\
J_{1_a1_b}&\simeq&-2J^{(1_A1_B)}_e\nonumber\\
J_{1_a2_b} &=& 0\nonumber\\
J_{2_a1_b} &=& 0\nonumber\\
J_{2_a2_b}&\simeq& -2J^{(2_A2_B)}_e,\nonumber
\end{eqnarray}
where
\begin{eqnarray}\label{deltab}
&\Delta E_{1_{a(b)}}=E_{(012,111)}(E_{(111,012)})-E_{(111,111)}\nonumber\\
&\Delta E_{2_{a(b)}}=E_{(102,111)}(E_{(111,102)})-E_{(111,111)}\nonumber\\
&\Delta E_{3_{a(b)}}=E_{(201,111)}(E_{(111,201)})-E_{(111,111)}\nonumber\\
&\Delta E_{4_{a(b)}}=E_{(021,111)}(E_{(111,021)})-E_{(111,111)}\nonumber
\end{eqnarray}
with the different terms defined in Eqs. \ref{b1}, \ref{b2}, \ref{b3}.
\begin{eqnarray}
\label{b1}
E_{(ijk,111)}&=& i\varepsilon_{1_a}+j\varepsilon_{2_a}+k\varepsilon_{3_a}+ijU_{1_a2_a}+ikU_{1_a3_a}+kjU_{2_a3_a}+\delta_{i2}U_{1_a}+\delta_{j2}U_{2_a}+ \delta_{k2}U_{3_a}+\nonumber\\
 &+&\varepsilon_{1_b}+\varepsilon_{2_b}+\varepsilon_{3_b}+U_{1_b2_b}+U_{1_b3_b}+U_{2_b3_b}+iU_{1_a1_b}+iU_{1_a2_b}+ jU_{2_a1_b}+jU_{2_a2_b}
\end{eqnarray}
\begin{eqnarray}\label{b2}
E_{(111,ijk)}&=&\varepsilon_{1_a}+\varepsilon_{2_a}+\varepsilon_{3_a}+U_{1_a2_a}+U_{1_a3_a}+U_{2_a3_a}+i\varepsilon_{1_b}+j\varepsilon_{2_b}+k\varepsilon_{3_b}+ ijU_{1_b2_b}+ikU_{1_b3_b}+\nonumber\\ 
&+& kjU_{2_b3_b}+\delta_{i2}U_{1_b}+\delta_{j2}U_{2_b}+\delta_{k2}U_{3_b}+iU_{1_a1_b}+ jU_{1_a2_b}+iU_{2_a1_b}+jU_{2_a2_b}
\end{eqnarray}
\begin{eqnarray}\label{b3}
E_{(111,111)}&=&\varepsilon_{1_a}+\varepsilon_{2_a}+\varepsilon_{3_a}+U_{1_a2_a}+U_{1_a3_a}+U_{2_a3_a}+\varepsilon_{1_b}+\varepsilon_{2_b}+\varepsilon_{3_b}+\nonumber\\
&+& U_{1_b2_b}+U_{1_b3_b}+U_{2_b3_b}+U_{1_a1_b}+U_{1_a2_b}+U_{2_a1_b}+U_{2_a2_b}
\end{eqnarray}

\bibliography{biblio.bib}